# How Silicon and Boron Dopants Govern the Cryogenic Scintillation Properties of *N*-type GaAs

S. Derenzo*[a] E. Bourret[a], C. Frank-Rotsch[b], S. Hanrahan[a], and M. Garcia-Sciveres[a]

[a]Lawrence Berkeley National Laboratory, 1 Cyclotron Road, Berkeley, CA 94720, USA
[b]Leibniz-Institut für Kristallzüchtung (IKZ), Max-Born-Str. 2, 12489 Berlin, Germany

**Abstract**

This paper is the first report describing how the concentrations of silicon and boron govern the cryogenic scintillation properties of *n*-type GaAs. It shows that valence band holes are promptly trapped on radiative centers and then combine radiatively with silicon donor band electrons at rates that increase with the density of free carriers. It also presents the range of silicon and boron concentrations needed for efficient light emission under X-ray excitation, which along with its low band gap and apparent absence of afterglow, make scintillating GaAs suitable for the detection of rare, low-energy electronic excitations from interacting dark matter particles. A total of 29 samples from four different suppliers were studied. Luminosities and timing responses were measured for the four principal emission bands centered at 860, 930, 1070, and 1335 nm, and for the total emissions. Excitation pulses of 40 kVp X-rays were provided by a light-excited X-ray tube driven by an ultra-fast laser. Scintillation emissions from 800 to 1350 nm were measured using an InGaAs photomultiplier. Within the concentration ranges of free carriers from 2 x $10^{16}$/cm$^3$ to 6 x $10^{17}$/cm$^3$ and boron from 1.5 x $10^{18}$/cm$^3$ to 6 x $10^{18}$/cm$^3$, nine samples have luminosities > 70 photons/keV and two have luminosities > 110 photons/keV. Other samples in that range have lower luminosities due to higher concentrations of non-radiative centers. The decay times decrease by typically a factor of ten with increasing free carrier concentrations from $10^{17}$/cm$^3$ to 2 x $10^{18}$/cm$^3$.

## 1. Introduction

In previous work, we determined that *n*-type GaAs doped with Si and B is a luminous cryogenic scintillator.[1, 2] In this work, we explore the dependence of its scintillation properties on the concentrations of Si and B. The paper is organized as follows: Section 2 describes 29 GaAs samples, the pulsed X-ray system, the filters used, and the calibration and analysis procedures. Section 3 presents the luminosities and time responses of the total emissions and of the four emission bands. Since the samples were provided after growth, it was not possible to control the doping levels. However, there are enough variations in doping levels to determine general trends. Section 3 also contains thermal quenching results for four samples that are bright in the four different emission bands. Section 4 lists the advantages for the direct detection of dark matter and Section 5 lists the conclusions. Table I defines the variables and abbreviations used. Appendix A contains tables of decay times and fractions of the four emission bands for the 29 samples and a plot of decay times vs. boron concentration.

Table I. Variables and abbreviations used in this paper

| | |
|---|---|
| AXT | AXT, Inc. (Fremont, CA) |
| BGO | Bismuth germanate scintillator ($Bi_4Ge_3O_{12}$) used for luminosity calibrations |
| $E_X$ | Photodetector quantum efficiency at the emission wavelengths of X = BGO, X = U (unfiltered average for GaAs), and GaAs emission band X = A, B, C, D |
| $f_m$ | Fraction of the $m^{th}$ exponential component |
| Frei | Freiberger, Inc. (Freiberg, Germany) |
| $f_{slow}$ | Fraction of light with decay times much longer than the 6 μs X-ray repetition period |
| GDMS | Glow Discharge Mass Spectrometry |
| $H$ | BGO luminosity calibration factor (equation 1) |
| IKZ | Leibniz-Institut für Kristallzüchtung, Berlin, Germany |
| $K_X$ | Weighting factors for converting ($P_X/Q_X$) values into luminosities for emission band X = A, B, C, D (equations 3–6) |
| $L_U$ | Total (unfiltered) luminosity (photons/keV) (equation 2) |
| $L_X$ | Luminosity of emission band X = A, B, C, D (photons/keV) (equations 3–6) |
| $M_e$ | Model of $N_e$ as a linear combination of $N_{Si}$ and $N_B$ |
| $N_e$ | N-type free carrier concentration in units of $10^{17}/cm^3$ |
| $N_Z$ | Concentration of element Z = Si, B, Al, P, S, Cl, Fe, Cu, Zn in units of $10^{17}/cm^3$ |
| $P_X$ | Photons detected during the acquisition time minus the phototube background (X = BGO, U, A, B, C, D, SP) |
| $Q_X$ | Average X-ray beam flux during the acquisition time (X = BGO, U, A, B, C, D, SP) |
| rms | standard deviation (root mean square of the deviations from the average) |
| SP | Subscript for short pass SP1200 filter values |
| U | Subscript for unfiltered values |
| UW | University Wafer, Inc. |
| $\tau_m$ | Decay time of the $m^{th}$ exponential component (ns) |

To summarize previous work, *N*-type GaAs(Si,B) has a Fermi level at the conduction band minimum and all electron traps are filled. It produces scintillation photons by the following steps: (1) an electronic excitation event leaves one or more holes in the valence band, (2) the valence band holes are trapped by acceptors and (3) the acceptor holes recombine radiatively with delocalized donor band electrons. Figure 1 shows the GaAs(Si,B) acceptor levels and emission energies previously identified [3-8]. Four radiative acceptors associated with luminescence have been identified as follows:

(A) Shallow defects [3-5]
(B) Boron on an arsenic site [3, 6]
(C) The silicon complex $Si_{Ga}V_{Ga}$ [7]
(D) The silicon complex $Si_{Ga}V_{Ga}Si_{Ga}$ [8]

In addition, a variety of deeper acceptors (X) have been identified [9, 10] that can trap valence band holes and reduce the luminosity below the theoretical limit of 220 photons/keV, a value based on the pair creation energy of 4.55 ± 0.02 eV at 0 K [11].

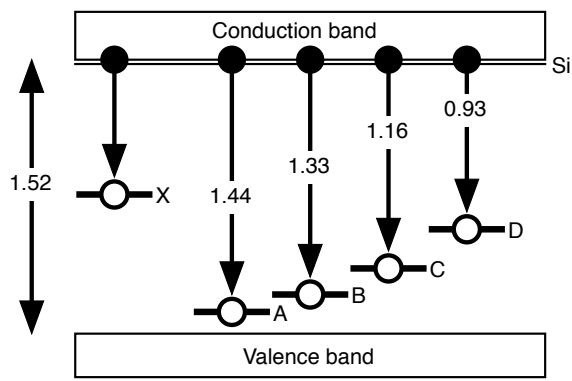

Figure 1. The GaAs energy gap and transition energies [2] (eV) for silicon donor electrons recombining with radiative acceptors A, B, C, and D, and with non-radiative acceptors X. Acceptors are shown empty as they would be after trapping a valence band hole.

## 2. Samples, equipment, and methods

### 2.1 Samples

Table II lists the 29 GaAs samples measured in this work, their suppliers, free carrier concentrations ($N_e$) and nine impurity concentrations. Elements heavier than Zn were below the limits of detectability (typically 0.001 to 0.005 ppm by weight) in all samples. Samples had typical areas of 10 mm x 10 mm and thicknesses varying from 0.3 to 2 mm, sufficient to absorb almost all the X-rays. The suppliers provided the free carrier concentration for each sample but did not know the silicon or boron concentrations. For this work the impurity levels for 70 elements from lithium to uranium were determined by GDMS (Evans Analytical Group, Santa Clara, CA). One sample (13354) was measured twice to check GDMS reproducibility.

### 2.2 Pulsed X-ray system

Figure 2 shows the pulsed X-ray system used in this work [12]. It consists of a cavity-dumped, frequency-doubled Ti-sapphire laser (Coherent, Inc., Santa Clara, CA) that drives a light-excited X-ray tube (N5084, Hamamatsu, Inc., Hamamatsu, Japan). Individual photons from the samples were recorded by an InGaAs phototube (R5509-43, Hamamatsu, Inc.) with a single photon time jitter of 1.5 ns and a quantum efficiency of 1–2% in the 300 to 1350 nm range. Optical emissions from the sample are made parallel by a quartz lens and focused on the 3 mm x 8 mm InGaAs photocathode by a second quartz lens. The different emission bands were selected by filters placed between the lenses. The lenses are transparent from 200 to 2200 nm. The phototube was calibrated by the manufacturer and the approximate quantum efficiencies for the various emission bands are $E_{BGO} = 1.1\%$, $E_A = 1.3\%$, $E_B = 2.1\%$, $E_C = 2.2\%$, and $E_D = 2.0\%$. The approximate average quantum efficiency over all GaAs emission bands is $E_U = 2\%$. Data analysis was simplified by the weak dependence of the quantum efficiency vs. wavelength over the GaAs emission spectrum.

Table II. 29 GaAs samples, their sources, $N_e$ and impurity levels (in units of $10^{17}/cm^3$). See Table I for terms used. Impurities other than Si and B above $10^{17}/cm^3$ are shown in boldface.

| Sample | Supplier | $N_e$ | $N_{Si}$ | $N_B$ | $N_{Al}$ | $N_P$ | $N_S$ | $N_{Cl}$ | $N_{Fe}$ | $N_{Cu}$ | $N_{Zn}$ |
|---|---|---|---|---|---|---|---|---|---|---|---|
| 13316 | AXT | 5.5 | 10.2 | 28.7 | 0.007 | 0.008 | 0.002 | <0.001 | <0.001 | <0.001 | <0.001 |
| 13330 | Frei | 6 | 9.47 | 15.4 | 0.166 | **2.90** | 0.030 | 0.253 | 0.017 | 0.020 | 0.001 |
| 13332 | Frei | 14.4 | 16.7 | 20.2 | 0.036 | **5.39** | 0.090 | 0.289 | 0.034 | 0.010 | 0.025 |
| 13333 | Frei | 20.3 | 22.8 | 23.4 | 0.024 | **4.55** | 0.020 | 0.072 | 0.006 | 0.003 | 0.010 |
| 13352 | AXT | 5.5 | 8.78 | 32.6 | 0.006 | 0.006 | 0.001 | <0.001 | <0.001 | <0.001 | <0.001 |
| 13353 | AXT | 5.5 | 8.21 | 27.3 | 0.006 | 0.003 | 0.001 | <0.001 | <0.001 | <0.001 | 0.001 |
| 13354[a] | AXT | 5.5 | 8.33 | 35.6 | 0.006 | 0.005 | 0.001 | <0.001 | <0.001 | <0.001 | <0.001 |
| 13354[a] | AXT | 5.5 | 9.24 | 38.5 | 0.036 | 0.003 | 0.020 | 0.018 | 0.006 | 0.005 | 0.015 |
| 13357 | IKZ | $3.8 \times 10^{-7}$ | 0.48 | 1.13 | 0.249 | 0.062 | 0.190 | **1.18** | 0.006 | 0.071 | 0.049 |
| 13358 | IKZ | 0.02 | 0.18 | 2.96 | 0.071 | 0.031 | 0.010 | 0.090 | 0.023 | 0.040 | 0.039 |
| 13359 | IKZ | 0.53 | 4.33 | 47.4 | 0.487 | 0.072 | 0.190 | 0.172 | 0.109 | 0.151 | 0.029 |
| 13363 | IKZ | 4.24 | 4.45 | 6.22 | 0.024 | <0.001 | 0.020 | 0.054 | 0.006 | 0.002 | 0.010 |
| 13364 | IKZ | 1.38 | 5.02 | 14.5 | **2.38** | 0.009 | 0.120 | 0.714 | 0.017 | 0.025 | 0.020 |
| 13365 | IKZ | 2.24 | 3.54 | 14.5 | **1.31** | 0.008 | 0.150 | 0.660 | 0.023 | 0.040 | 0.044 |
| 13701[b] | IKZ | 4.6 | 9.24 | 54.6 | 0.001 | 0.012 | 0.003 | 0.001 | <0.001 | <0.001 | 0.001 |
| 13702 | IKZ | 5.1 | 7.76 | 44.5 | 0.332 | 0.010 | 0.020 | 0.361 | 0.052 | 0.003 | 0.005 |
| 13703 | IKZ | 7 | 8.90 | 41.5 | 0.214 | 0.021 | 0.040 | **1.27** | <0.003 | <0.001 | <0.001 |
| 13704.1 | IKZ | 4.3 | 5.02 | 13.0 | 0.024 | 0.003 | 0.140 | 0.262 | <0.003 | <0.001 | 0.044 |
| 13705.2 | IKZ | 5.8 | 6.96 | 19.3 | 0.012 | <0.001 | 0.210 | 0.045 | <0.003 | <0.001 | <0.001 |
| 13706 | IKZ | 6 | 17.1 | 53.3 | 0.772 | <0.001 | 0.240 | 0.208 | 0.143 | 0.136 | 0.039 |
| 13707 | IKZ | 0.23 | 1.25 | 35.6 | 0.071 | 0.021 | 0.020 | 0.090 | <0.003 | 0.020 | 0.005 |
| 13708 | IKZ | 0.52 | 2.74 | 38.5 | 0.036 | 0.003 | 0.040 | 0.081 | <0.003 | 0.010 | 0.010 |
| 13718[d] | UW | 2.7[c] | 4.22 | 21.0 | 0.048 | <0.001 | 0.03 | 0.027 | 0.006 | <0.001 | 0.005 |
| 13720[e] | UW | 1.2[c] | 2.17 | 16.9 | 0.392 | <0.001 | 0.02 | 0.235 | 0.004 | <0.001 | 0.004 |
| 40017 | IKZ | 7.6 | 9.70 | 32.6 | 0.416 | 0.010 | 0.050 | 0.280 | 0.011 | 0.020 | 0.034 |
| 40018 | IKZ | 5.6 | 8.44 | 56.3 | 0.214 | 0.021 | 0.020 | 0.072 | 0.017 | 0.015 | 0.010 |
| 40019 | IKZ | 6.8 | 9.81 | 32.6 | 0.071 | 0.021 | 0.130 | 0.361 | 0.023 | 0.020 | 0.010 |
| 40020 | IKZ | 5.2 | 10.3 | 44.5 | **5.94** | <0.001 | 0.160 | 0.163 | 0.029 | 0.015 | 0.304 |
| 40021 | IKZ | 9 | 20.5 | 32.6 | 0.059 | 0.004 | 0.050 | 0.172 | 0.029 | 0.071 | **1.176** |
| 40022 | IKZ | 7.6 | 11.4 | 38.5 | 0.024 | 0.010 | 0.040 | 0.072 | 0.011 | 0.005 | 0.348 |

[a] Duplicate GDMS measurements
[b] GDMS measured by CMK Ltd., Zarnovica, Slovakia, all others by EAG (Evans Analytical Group, Santa Clara, CA)
[c] Estimated from $N_{Si}$, $N_B$ and equation 7
[d] 13718 (this table) and 13719 (table V) are neighboring cuts from the top of a 300 g crystal
[e] 13720 (this table) and 13721 (table V) are neighboring cuts from the bottom end of the same 300 g crystal

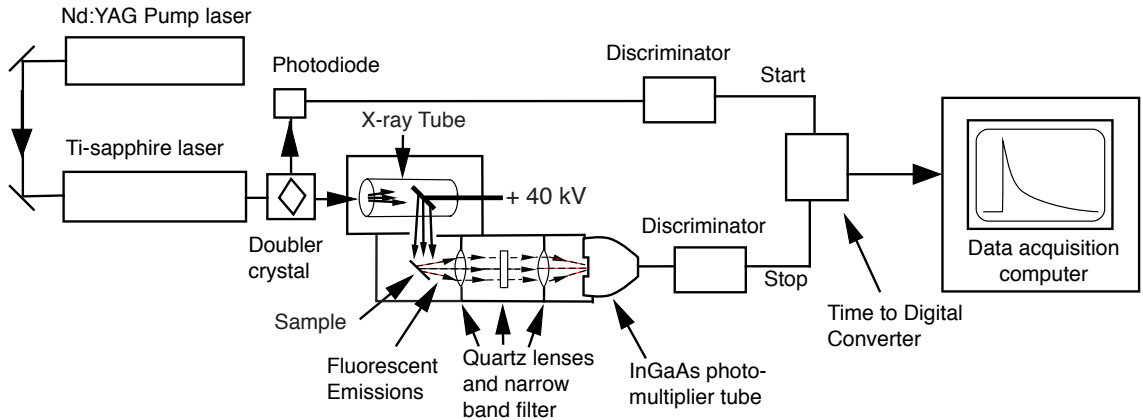

Figure 2. Experimental setup for the pulsed X-ray measurements reported in this paper.

**2.3 Filters used to sample emission bands A, B, C, D**

The emission spectra of the four principal emission bands of GaAs(Si,B) have been previously reported.[2] Table III lists the filters (Edmund Optics Inc., Cupertino, CA), and Table IV lists the filter combinations used to measure the luminosities and decay curves of the four emission bands. The SP1200 filter was used to block band D emissions that are beyond the stopbands of the NB850, NB925, and NB1075 filters. According to the Edmund Optics datasheets for these filters, transmission is typically 99% in the passbands and $10^{-5}$ in the stopbands.

Table III Filters used in this work.

| Edmund Optics catalog # | Filter type | Filter passband (nm) | Filter stopband (nm) |
|---|---|---|---|
| 84-802 | Narrow Band NB850 | 825–875 | 875–1200 |
| 86-971 | Narrow Band NB925 | 900–950 | 950–1200 |
| 87-874 | Narrow Band NB1075 | 1050–1100 | 1100–1500 |
| 85-911 | Narrow Band NB1350 | 1325–1375 | 1375–1500 |
| 89-678 | Short Pass SP1200 | 200–1200 | 1200–1650 |

Table IV Filter combinations for measuring emission bands A, B, C, and D.

| Emission band | Peak wavelength [2] (nm) | Filter combination |
|---|---|---|
| A | 860 | NB850 + SP1200 |
| B | 930 | NB925 + SP1200 |
| C | 1070 | NB1075 + SP1200 |
| D | 1335 | NB1350 |

**2.4 Luminosity calibrations**

Before measuring the luminosity and time response of each GaAs sample, a BGO crystal of the same size was placed in the cryostat and the number of detected photons minus the phototube background ($P_{BGO}$) and the average X-ray beam flux ($Q_{BGO}$) were recorded for 1200 s. This calibration was done at room temperature, where the luminosity of BGO (8.2 photons/keV) is well known [13]. The GaAs sample was then placed in the cryostat with a gold reflector backing, cooled to 10 K, and the corresponding unfiltered values $P_U$ and $Q_U$ were recorded over 1200 s.

The BGO calibration factor is given by:
$$H = (P_{BGO}/Q_{BGO})/E_{BGO} \qquad (1)$$

For each sample, the total luminosity is given by:
$$L_U = 8.2\,(P_U/Q_U)/(H\,E_U) \qquad (2)$$

The excitation rate was 165 kHz and the X-ray flux was adjusted by automatically selecting neutral density filters in the laser beam to keep the detected photon/excitation ratio < 0.02. Scintillation photons were recorded at a typical rate of 3000/s during a 1200 s acquisition period. In operation the phototube is cooled to –80 C and has a dark pulse background rate of about 50/s. To measure the rate of background pulses, data were collected for 120 s before and after each 1200 s data run. No antireflection coatings were applied.

To estimate the luminosities and decay times of the four emission bands, the four filter combinations listed in Table IV were placed between the sample and the phototube. The number of photons minus the phototube background ($P_X$) and the average X-ray flux ($Q_X$) were measured over a 1200 s acquisition period, where X = A, B, C, and D. The luminosities of the four emission bands are determined using the following equations:

$L_A = 8.2\,(K_A P_A/Q_A)/(H\,E_A)$  (3)
$L_B = 8.2\,(K_B P_B/Q_B)/(H\,E_B)$  (4)
$L_C = 8.2\,(K_C P_C/Q_C)/(H\,E_C)$  (5)
$L_D = 8.2\,(K_D P_D/Q_D)/(H\,E_D)$  (6)

The $K_A$, $K_B$, $K_C$, and $K_D$ are weighting factors that convert the narrow band $P_X/Q_X$ values into the full emission band luminosities in photons/keV. They were determined by a least-squares fit of the sum of the weighted filtered values $K_A P_A/Q_A + K_B P_B/Q_B + K_C P_C/Q_C + K_D P_D/Q_D$ to the unfiltered values $P_U/Q_U$ for the 29 samples. The best-fit values are $K_A = 1.1$, $K_B = 2.1$, $K_C = 2.8$, and $K_D = 3.4$. As a check, the value of $K_D$ agrees with the ratio $(P_U/Q_U - P_{SP}/Q_{SP})/(P_D/Q_D)$. The determination of the weighting factors from the data is possible because different samples have different values of $P_X/Q_X$ in the four emission bands.

## 2.5 Fitting decay times and fractions to the decay curves

Reference [12] describes the analysis used to determine the minimum number of exponential components needed to fit the decay curves while avoiding local minima.

## 3. Results and discussion

### 3.1 Silicon-boron charge distribution

In traditional semiconductors the *n*-type free carrier concentration is simply the excess of the donor concentration above the acceptor concentration. GaAs(Si,B) is more complicated. Table II shows that although all samples are *n*-type, $N_B$ is higher than $N_e$, and in some cases considerably higher. This is in agreement with evidence that most of the boron resides on isoelectronic gallium sites and that the boron acceptor emission arises from a smaller fraction on arsenic acceptor sites [14].

In this work we relate the free carrier concentration to a linear combination of $N_{Si}$ and $N_B$:
$M_e = a N_{Si} - b N_B$   (7)

Using the $N_e$, $N_{Si}$, and $N_B$ values in Table II, the best least-squares parameters in the model $M_e$ fitted to the $N_e$ values are $a = 0.795$ and $b = 0.032$ (Figure 3). According to this fit, about 76% (calculated as $a - b$) of the silicon donor atoms provide free carriers and 3.2% provide electrons to the boron acceptors. Since the material is *n*-type and all acceptor levels are filled, there may be other boron atoms on arsenic sites that received electrons from the dissociation of the $B_2O_3$ flux used during crystal growth.

Both samples 40021 and 13706 are outliers in that they have abnormally low $N_e$ values and were not included in the fit. Sample 40021 is unusual in that it has a relatively high zinc content. Sample 13706 is unusual in that it has abnormally low luminosities in all emission bands as well as a low value of $N_e$, both possibly due to relatively high levels of electron trapping defects.

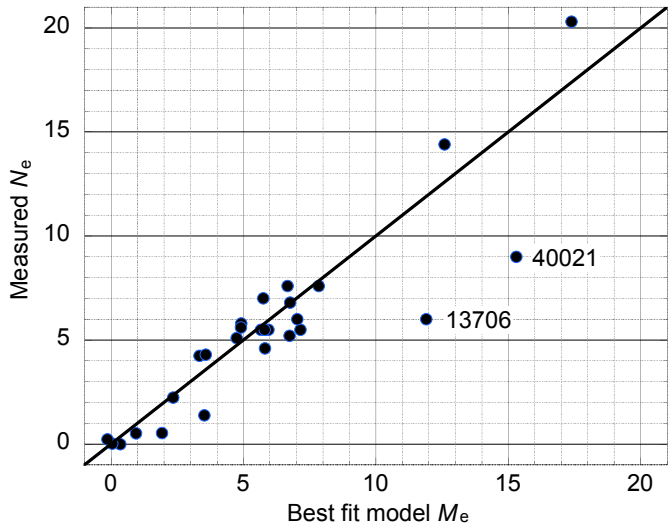

Figure 3. $N_e$ vs. the best-fit values of $M_e$ (equation 7). Outlier samples 13706 and 40021 were not included in the fit.

## 3.2 Luminosities vs. $N_e$ and $N_B$

Table V lists the 29 samples, their $N_e$, $N_{Si}$, and $N_B$ values, the total (unfiltered) luminosities $L_U$, and the luminosities in emission bands A, B, C, and D (calculated from equations 3–6), sorted by $L_U$.

Considering the high refractive index of GaAs (about 3.4 at 1000 nm),[15] the high luminosities without antireflection coatings is surprising. This issue needs to be explored in future work, especially the absorption and scattering of photons by the free carriers. If necessary, antireflection coatings can be used to reduce internal trapping.[16]

Note that the quantum efficiency of the R5509-43 phototube decreases above 1350 nm, which is near the peak of emission band D. Since band D extends to 1600 nm [2] a large fraction (possibly 50%) of the band D emissions are not included in $L_D$ or $L_U$. Gamma ray spectroscopy would be a better way to measure luminosities, using a cryogenic germanium photodiode with a high quantum efficiency over the full GaAs emission spectrum.

### 3.2.1 Total luminosities

Figure 4 shows a scatter plot of $L_U$ vs. $N_e$ and $L_U$ vs. $N_B$, using data from Table V. The nine samples with values of $L_U$ above 70 occur over a wide range of $N_e$, from 0.23 to 6 and $N_B$ from 15 to 55. However, low values of $L_U$ also occur in this range due to the presence of nonradiative acceptors (X in figure 1). Samples 13707 and 13359 have high total luminosities but some of the lowest $N_e$ values (0.23 and 0.53, respectively). Apparently, an $N_e$ value as low as 0.23 is sufficient for high luminosity. Low values of $N_e$ may be desirable if optical absorption by free carriers becomes important in large crystals. Values of Ne above 10 may be evidence of Auger quenching, where a donor electron recombining with a hole imparts its energy to another donor electron rather than producing a photon. The two samples with the highest values of $L_U$ have relatively high $N_B$ values. The three samples with $N_B < 10$ have low values of $L_U$.

Table V. Sample numbers, suppliers, $N_e$, $N_{Si}$, $N_B$, and luminosities for the 29 samples at 10 K. Sorted by total luminosity $L_U$ (in boldface).

| Sample | Supplier | $N_e$ | $N_{Si}$ | $N_B$ | $L_U$ | $L_A$ | $L_B$ | $L_C$ | $L_D$ |
|---|---|---|---|---|---|---|---|---|---|
| 13357 | IKZ | 3.8x10$^{-7}$ | 0.48 | 1.13 | **0.30** | 0.22 | 0.01 | 0.01 | 0.06 |
| 13358 | IKZ | 0.02 | 0.18 | 2.96 | **5.00** | 2.07 | 0.20 | 0.71 | 2.04 |
| 13706 | IKZ | 6 | 17.1 | 53.3 | **11.3** | 1.96 | 3.95 | 4.46 | 0.96 |
| 13333 | Frei | 20.3 | 22.8 | 23.4 | **13.3** | 5.98 | 3.70 | 2.79 | 0.87 |
| 13703 | IKZ | 7 | 8.90 | 41.5 | **17.0** | 1.22 | 10.4 | 3.61 | 1.67 |
| 40021 | IKZ | 9 | 20.5 | 32.6 | **17.3** | 2.39 | 5.16 | 8.81 | 0.96 |
| 40018 | IKZ | 5.6 | 8.44 | 56.3 | **21.0** | 2.59 | 14.1 | 2.73 | 1.62 |
| 40017 | IKZ | 7.6 | 9.70 | 32.6 | **22.3** | 4.43 | 11.8 | 3.91 | 2.18 |
| 40022 | IKZ | 7.6 | 11.4 | 38.5 | **22.9** | 1.23 | 9.89 | 8.39 | 3.38 |
| 13363 | IKZ | 4.24 | 4.45 | 6.22 | **26.4** | 0.13 | 13.7 | 5.82 | 6.78 |
| 40019 | IKZ | 6.8 | 9.81 | 32.6 | **28.6** | 3.70 | 18.4 | 4.74 | 1.74 |
| 13332 | Frei | 14.4 | 16.7 | 20.2 | **32.6** | 5.45 | 5.46 | 19.67 | 2.04 |
| 13702 | IKZ | 5.1 | 7.76 | 44.5 | **39.3** | 7.91 | 26.3 | 3.21 | 1.86 |
| 13365 | IKZ | 2.24 | 3.54 | 14.5 | **39.5** | 2.24 | 8.40 | 4.89 | 24.0 |
| 40020 | IKZ | 5.2 | 10.3 | 44.5 | **42.5** | 3.55 | 27.8 | 7.18 | 3.96 |
| 13704.1 | IKZ | 4.3 | 5.02 | 13.0 | **43.9** | 2.93 | 22.1 | 8.77 | 10.1 |
| 13705.2 | IKZ | 5.8 | 6.96 | 19.3 | **48.5** | 5.96 | 33.3 | 5.28 | 3.94 |
| 13330 | Frei | 6 | 9.47 | 15.4 | **55.9** | 1.29 | 8.74 | 29.41 | 16.5 |
| 13708 | IKZ | 0.52 | 2.74 | 38.5 | **57.4** | 0.85 | 26.0 | 4.61 | 25.9 |
| 13721 | UW | 1.18 | 2.17 | 16.9 | **61.6** | 0.57 | 0.36 | 8.96 | 51.7 |
| 13719 | UW | 2.68 | 4.22 | 21.0 | **72.8** | 1.17 | 0.73 | 26.94 | 43.9 |
| 13364 | IKZ | 1.38 | 5.02 | 14.5 | **79.6** | 0.52 | 4.24 | 7.05 | 67.7 |
| 13707 | IKZ | 0.23 | 1.25 | 35.6 | **80.6** | 1.77 | 49.1 | 4.86 | 24.8 |
| 13316 | AXT | 5.5 | 10.2 | 28.7 | **83.1*** | 1.32 | 26.4 | 25.83 | 29.6 |
| 13353 | AXT | 5.5 | 8.21 | 27.3 | **84.2** | 0.95 | 24.8 | 27.50 | 31.0 |
| 13354 | AXT | 5.5 | 8.79 | 37.1 | **93.4** | 1.02 | 24.6 | 31.08 | 36.7 |
| 13701 | IKZ | 4.6 | 9.24 | 54.6 | **96.2** | 18.1 | 69.0 | 5.44 | 3.60 |
| 13352 | AXT | 5.5 | 8.78 | 32.6 | **117.5** | 1.42 | 31.4 | 37.41 | 47.3 |
| 13359 | IKZ | 0.53 | 4.33 | 47.4 | **118.3** | 2.89 | 94.9 | 4.92 | 15.6 |

* Good agreement with 67 ph/keV in [2]

Figure 5 explores possible correlations between $L_U$, $N_e$, and $N_B$ by plotting each sample according to its $N_e$ and $N_B$ value as a circle whose size is related to its luminosity. The nine samples with $L_U > 70$ are distributed between values of $N_e$ between 0.23 and 5.5 and $N_B$ between 14.5 and 55. There is no apparent clustering within these bounds. The low values of both $N_B$ and $N_e$ in the lower left-hand corner (samples 13357 and 13358) apparently do not provide sufficient concentrations of radiative acceptors and free carriers for good luminosity. These data show that high luminosities are possible over a range of values of $N_e$ and $N_B$.

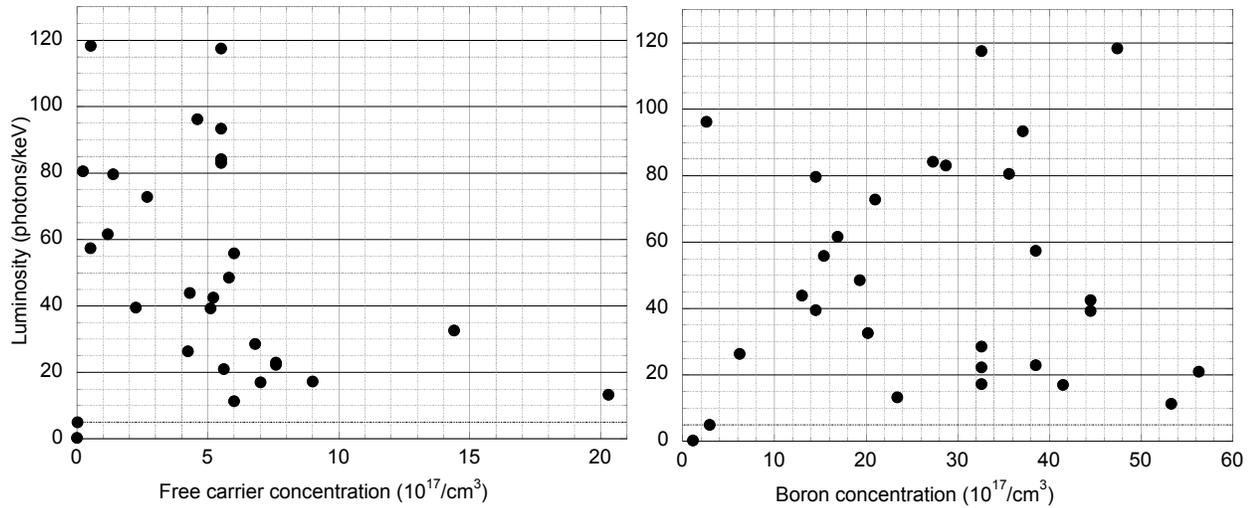

Figure 4. Scatter plot of $L_U$ vs $N_e$ and $L_U$ vs $N_B$ for the 29 samples.

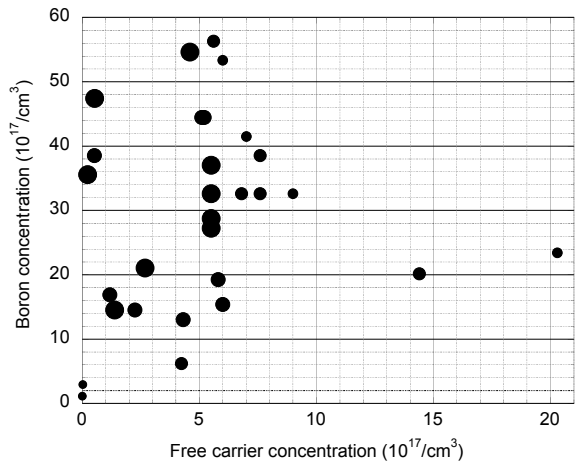

Figure 5. Scatter plot of the 29 samples plotted according to their $N_e$ and $N_B$ values. Ascending circle sizes correspond to $L_U$ <10, 10-20, 20-35, 35-70, and 70-118.

### 3.2.2 Luminosities for bands A, B, C, and D.

The previous section described the relation between $L_U$, $N_e$, and $N_B$. Figure 6 explores the same relations for the four emission bands that contribute to $L_U$.

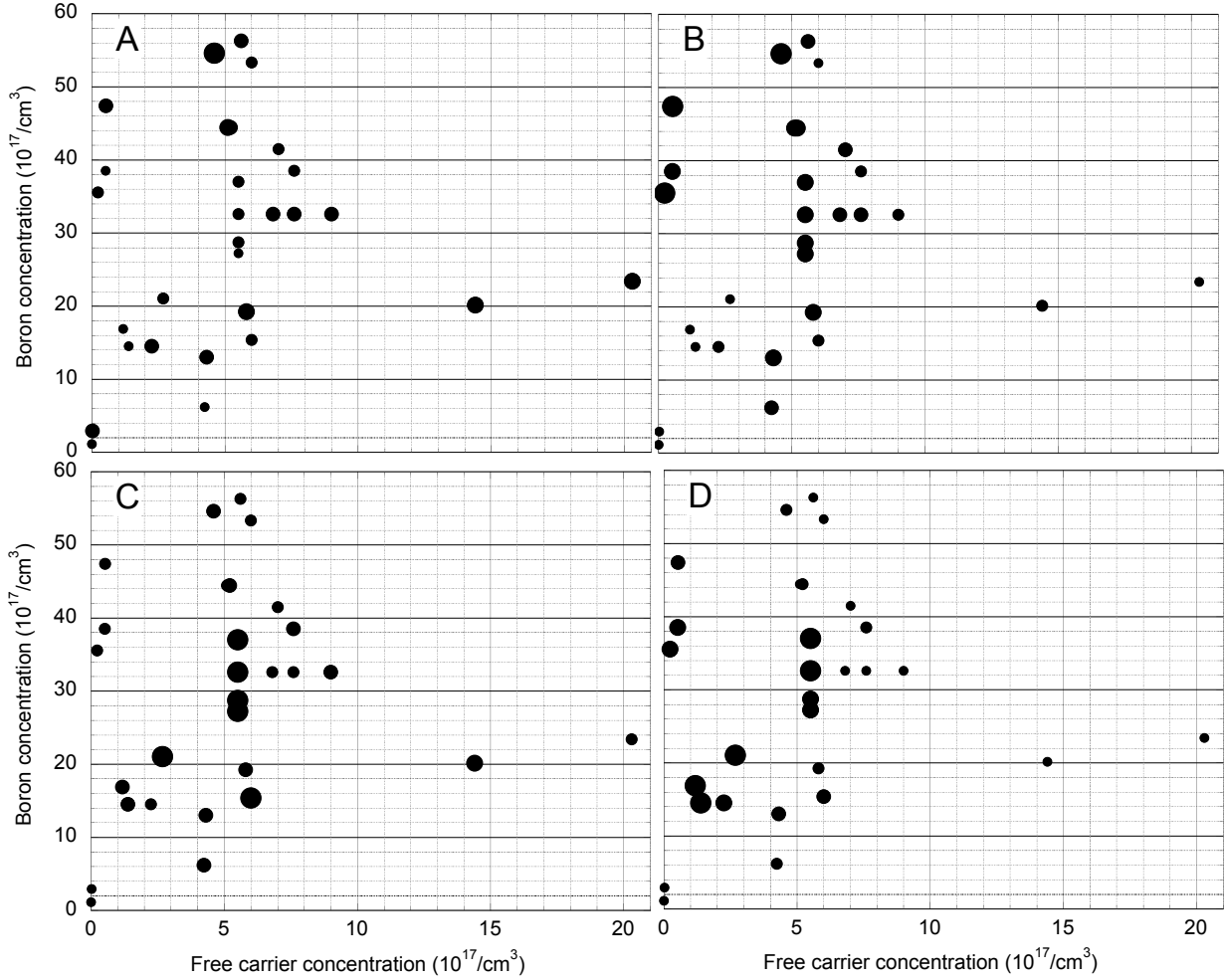

Figure 6. Scatter plots of the 29 samples in the four emission bands A, B, C, and D plotted according to their $N_e$ and $N_B$ values. Ascending circle sizes in plot A correspond to $L_A$ <1, 1-2, 2-5, 5-10, and 10-18. In plot B these are $L_B$ <5, 5-10, 10-20, 20-40, and 40-95. In plot C these are $L_C$ <2, 2-5, 5-10, 10-38. In plot D these are LD <3, 3-10, 10-20, 20-35, and 35-68.

For emission band A, sample 13701 has an $L_A$ (18) that exceeds that of the other samples by a large factor. This is surprising in view of the large value of $N_B$. Samples 13357 and 13358 have low luminosities in all bands, as expected from the low values of $N_e$, $N_{Si}$, and $N_B$. However, the fraction of their light in emission band A is much higher than the other samples. This agrees with the association of band A with shallow defects that are not associated with boron or silicon.

For emission band B, sample 13359 has the highest $L_U$ (118), the highest $L_B$ (95), and among the highest values of $N_B$ (47). It is the best example of boron acceptor luminescence into emission band B and evidence that a relatively low value of $N_e$ (0.53) is sufficient for high luminosity. Sample 13701 has the second-highest $L_B$ (69) and also a high value of $N_B$ (54.6).

For emission band C, six samples have $L_C$ between 26 and 37, $N_e$ between 2.7 and 6 and $N_B$ between 4.4 and 10.

For emission band D, sample 13364 has the highest $L_D$ (67.7) and much lower than average luminosities in all the other bands. Although band D has been associated with silicon complexes, there is little correlation between $L_D$ and $N_{Si}$.

## 3.3 Multi-exponential fits to the band A, B, C, and D emissions

This section presents multi-exponential fits to the band A, B, C, and D decay curves, as measured using the filter combinations listed in Table IV. Each filter has a 50 nm wide bandpass centered near the peak of the corresponding emission band. This provides decay curves that are most characteristic of their bands. All fits need a "slow" component whose decay time is significantly longer than the 6 μs pulsed X-ray repetition rate. Most of the samples have sufficiently high decay curve statistics to require a rise time, and the average value is 0.8 ns (rms 0.2 ns), consistent with the transit time spread of the phototube. The uncertainties in the fitted values of the decay times and fractions were estimated by measuring sample 13702 five times on different days over a 12-month period. The rms of the decay times is 9% of their values, and the rms of the fractions is 0.05.

The radiative states responsible for the different emission bands do not have well-defined decay times. For all emission bands the decay times decrease typically by about a factor of 10 as $N_e$ is increased from 0.2 to 20. For example, sample 13333 has the highest value of $N_e$ (20.3) and the shortest decay times for all components and all emission bands. Samples with $N_e < 1$ have the longest decay times for all emission bands.

Figure 7 shows the exponential component decay times and Figure 8 shows the fractions for emission bands A, B, C, and D as a function of $N_e$. In addition to a slow component, bands A, B, and C required two exponential components and band D required one component. Tables listing the luminosities, decay times and fractions of the four emission bands from each sample are provided in Appendix A. We stress that the decay times and fractions are provided only as analytical descriptions of the decay curves and are not the characteristics of well-defined radiative states

This trend of decreasing decay times with increasing $N_e$ (Figure 7) is evidence that GaAs(Si,B) is not a classic scintillator with a decay time characteristic of a self-trapped exciton (e.g. $BaF_2$ and CsI) or a luminous ion (e.g. $Ce^{3+}$), but that its decay time is controlled by the concentration of free electrons in the vicinity of acceptors that have trapped a valence band hole.

Figure 1 in Appendix A is a plot of the seven decay times as a function of $N_B$. It shows that unlike the dependence on $N_e$ shown in Figure 7, none of the decay times have a systematic dependence on $N_B$ from $N_B = 6$ to 55.

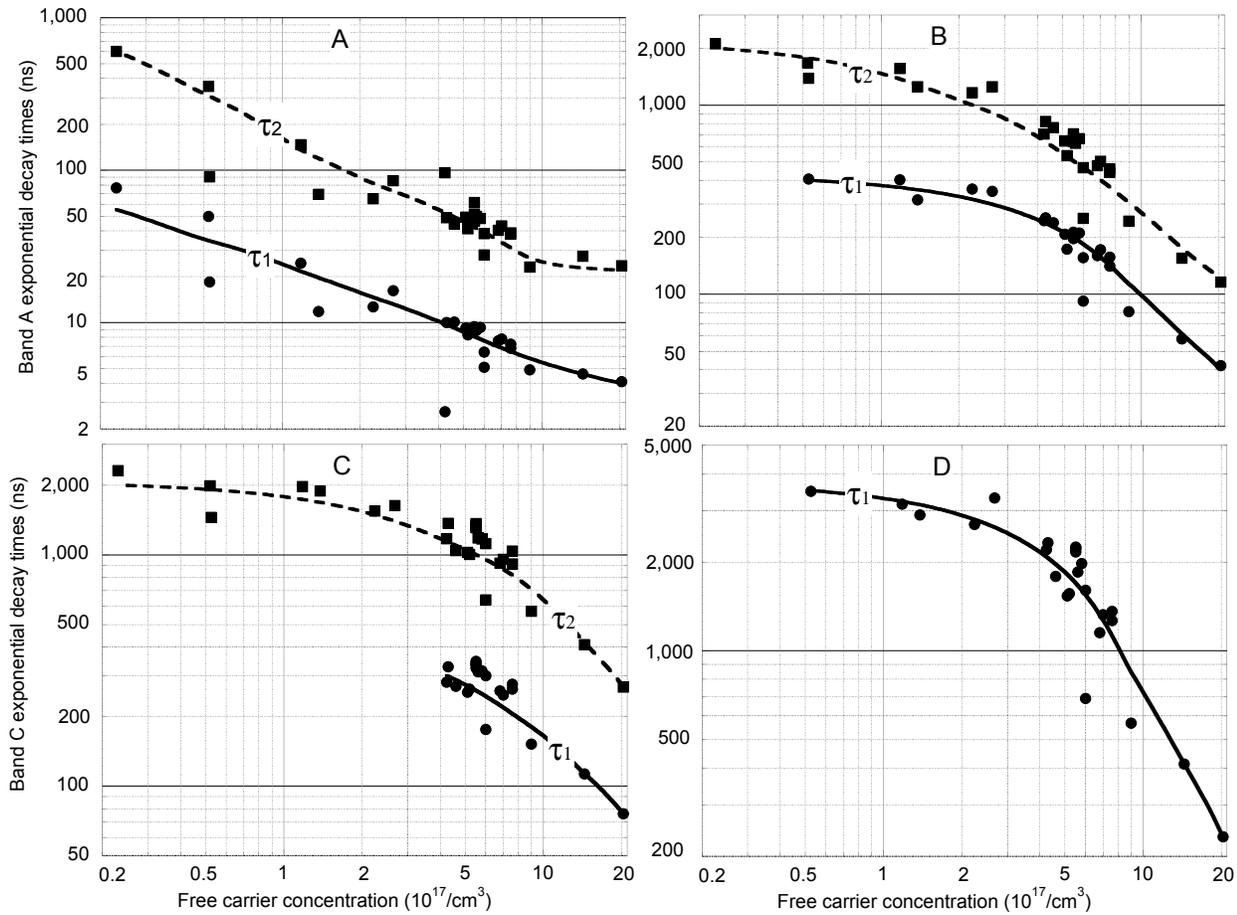

Figure 7. $\tau_1$ (circles) and $\tau_2$ (squares) for emission bands A, B, C, and D as a function of $N_e$. Only components with $N_e > 0.02$ and fractions $>0.05$ have been included. Trend lines are provided to guide the eye through the sample-to-sample variations.

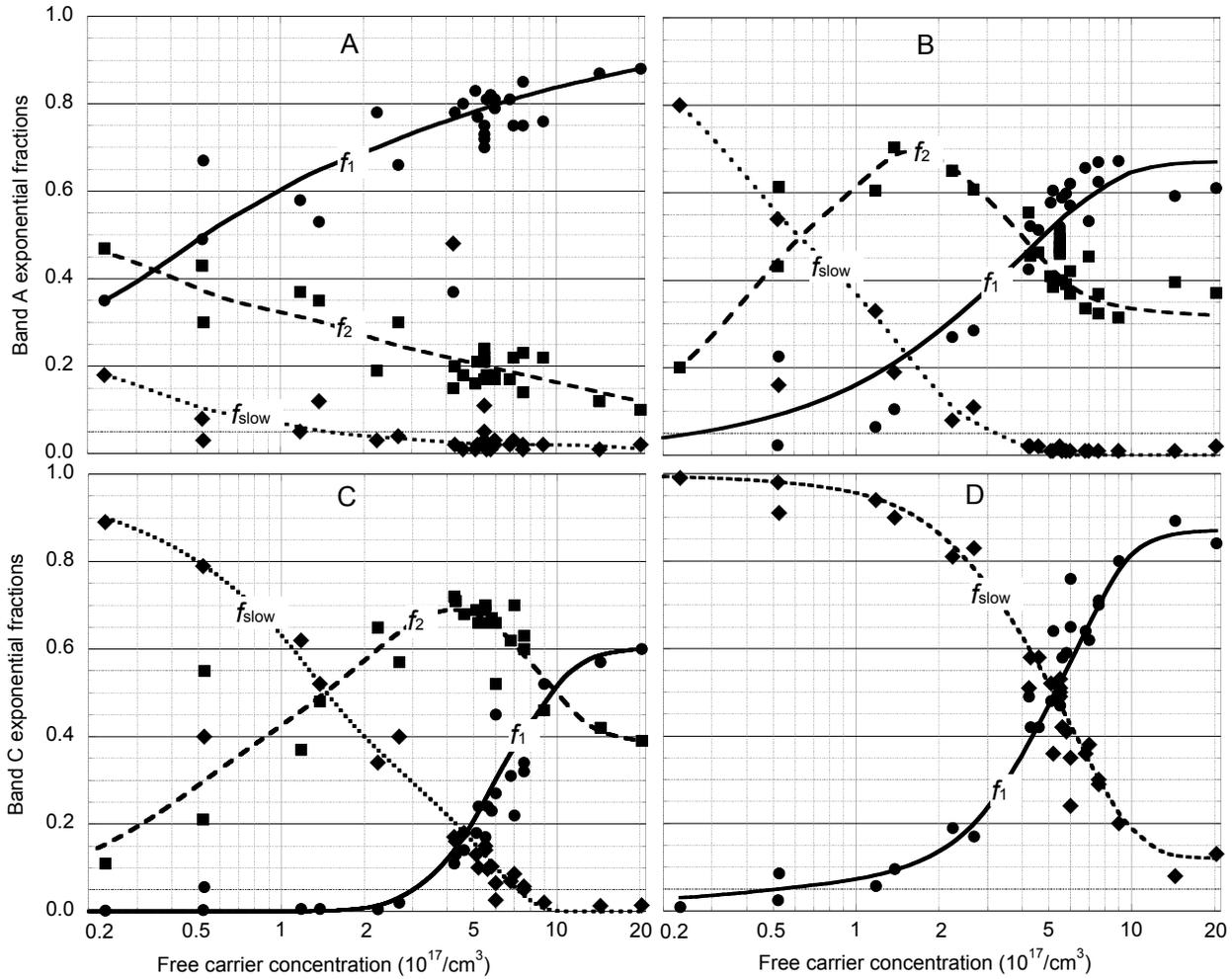

Figure 8. $f_1$ (circles), $f_2$ (squares) and $f_{slow}$ (diamonds) for emission bands A, B, C, and D as a function of $N_e$. Trend lines are provided to guide the eye through the sample-to-sample variations.

**3.4 Thermal quenching**

Figure 9 shows the thermal quenching of the luminosities of the four emission bands. Four samples were selected that have particularly strong emissions in each band. Sample 13701 was chosen for emission band A, sample 13359 for emission band B, 13330 for emission band C, and sample 13364 for emission band D. Data were taken under the same conditions as for Figures 7 and 8. The luminosities of the emission bands are quenched at different rates by increasing temperature. Only band D has significant luminosity at room temperature.

Figures 10 and 11 show that increasing temperature does not appreciably change the decay times or fractions for the seven exponential components that describe the decay curves of bands A, B, C, and D. This plus the sub-ns rise time are evidence that thermal quenching only reduces the prompt trapping of valence band holes on radiative acceptors and does not cause depopulation of the radiative states during emission. This is different from the usual thermal quenching where excited states are depopulated after excitation and the decay time and luminosity both decrease with increasing temperature. For samples that have a high luminosity at low temperatures, a large fraction of the valence band holes are trapped on the radiative centers. At sufficiently high temperatures, almost all the valence band holes are trapped on non-radiative centers and the luminosity approaches zero. This is consistent with the model that valence band holes are promptly trapped on radiative and non-radiative acceptors in a ratio that is temperature dependent.

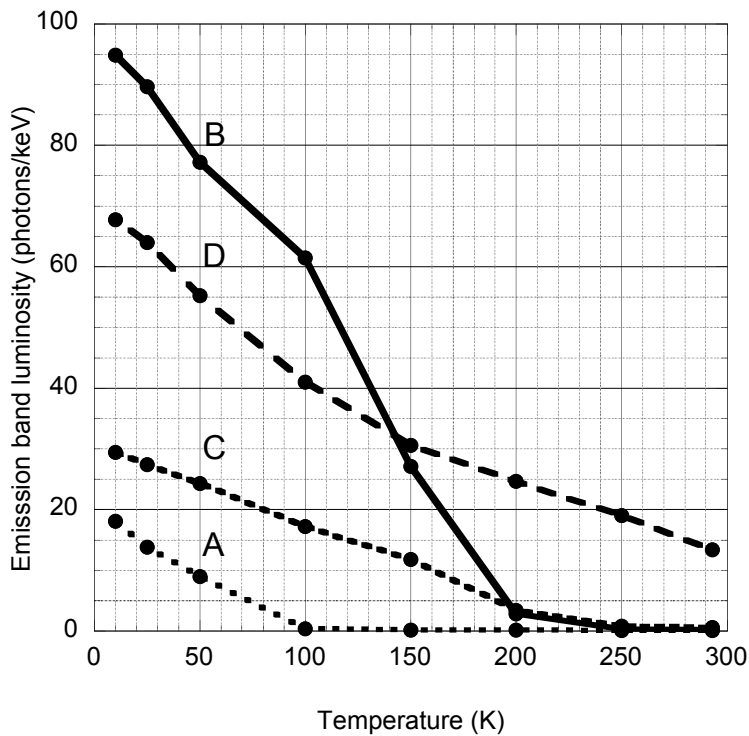

Figure 9. Emission band A, B, C, and D luminosities vs. temperature.

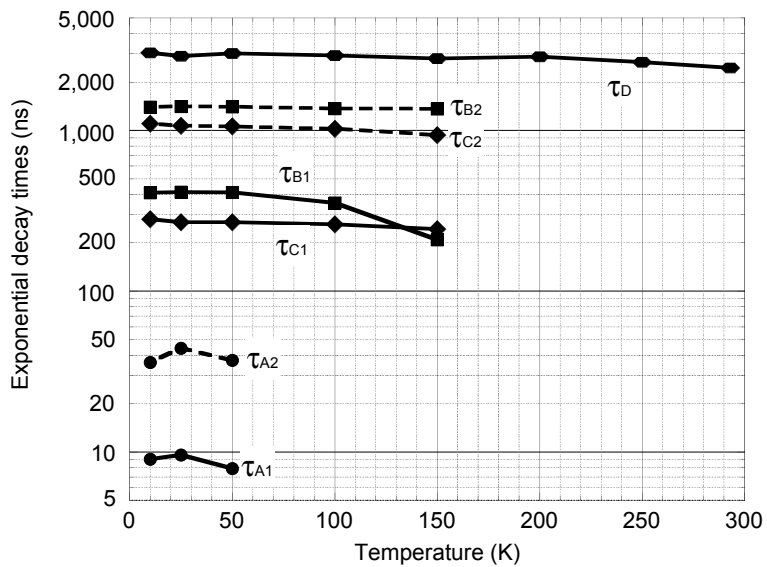

Figure 10. Exponential decay times for the emission bands A (circles), B (squares), C (diamonds), and D (rectangles) as a function of temperature. Solid lines for component 1, dashed lines for component 2. Data for bands A, B and C were not plotted at the higher temperatures due to low statistics. The decay times are essentially independent of temperature over the full quenching range, indicating that thermal quenching does not depopulate the radiative states.

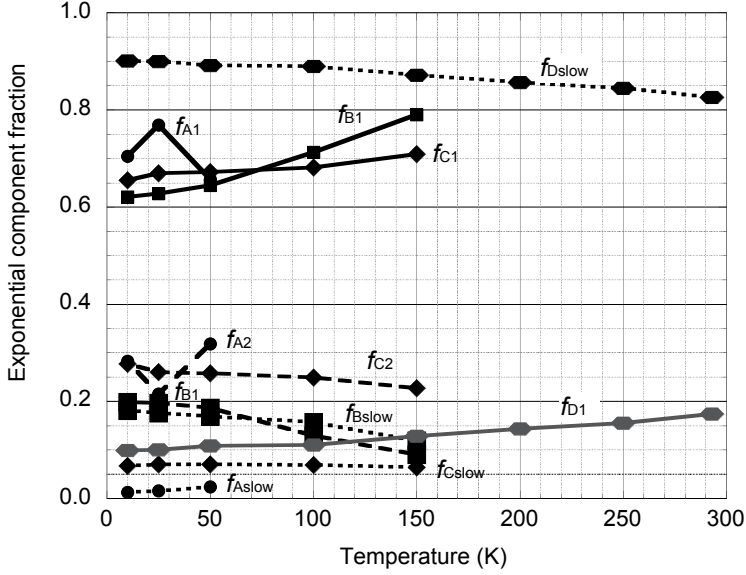

Figure 11. Fractions of the exponential components for the emission bands A (circles), B (squares), C (diamonds), and D (rectangles) as a function of temperature. Solid lines for $f_1$, long dashed lines for $f_2$, short dashed lines for $f_{slow}$.

## 4. Detection of Interacting Dark Matter

Scintillating GaAs is an excellent target for the detection of rare, low-energy electronic excitations from interacting dark matter particles for the following reasons:

- There are no naturally occurring radioactive isotopes of Ga or As.
- The band gap is 1.52 eV, making possible the detection of electronic excitations from interacting dark matter particles as light as a few MeV/$c^2$ for electron elastic scattering [17] and a few eV/$c^2$ for inelastic processes [18].
- *N*-type GaAs is commercially grown as 5 kg crystals.
- The donor electrons do not freeze out above a free-carrier concentration of 8 x $10^{15}$/cm$^3$ [19].
- Metastable radiative states that could cause afterglow are annihilated by delocalized donor electrons, as evidenced by the lack of thermally stimulated luminescence [1].
- Luminosities >110 photons/keV are observed without antireflection coatings (this paper).

A key challenge for the application of scintillating GaAs in the detection of interacting dark matter is the development of large-area cryogenic infra-red photon detectors that have very low dark pulse rates. Three candidates for development are transition edge sensors [20], superconducting nanowire single photon detectors, and microwave kinetic induction detectors.

# 5. Conclusions

In summary, we measured the luminosities and decay times of 29 GaAs samples that have different concentrations of silicon and boron. Nine have scintillation luminosities above 70 photons/keV and two are above 115 photons/keV. The data allow us to make the following conclusions:

- The highest-luminosity samples have free carrier concentrations from $2 \times 10^{16}$/cm$^3$ to $6 \times 10^{17}$/cm$^3$ and boron concentrations from $1.5 \times 10^{18}$/cm$^3$ to $6 \times 10^{18}$/cm$^3$, so precise dopant control during crystal growth may not be necessary in practical applications. However, control of non-radiative centers will be necessary for consistently high luminosity.
- The rise times for all emission bands is about 0.8 ns and limited by system response, so valence band holes are trapped promptly by the radiative acceptors.
- The emission band decay times decrease by about a factor of ten as the sample free carrier concentrations increase from $10^{17}$/cm$^3$ to $2 \times 10^{18}$/cm$^3$, so the recombination rates depend almost linearly on the concentration of free carriers. However, none of the decay times have a systematic dependence on the boron concentration.
- The decay times differ from band to band in the same sample, so the recombination rates depend on the electronic structures of the different acceptors and their interaction with free carriers.
- Thermal quenching decreases the luminosities of all four emission bands but has little effect on their decay times, so (1) the quenching is due to temperature-dependent differences in hole trapping by radiative and non-radiative acceptors and (2) once they have trapped a hole, the radiative acceptors are thermally stable.

**Acknowledgements**

We thank R. Essig for pointing out the need for a dark matter scintillator target with a low band gap, M. Boswell, L. Brown, E. Hernandez, F. Moretti, and D. Onken for assistance in data acquisition, and M. Pyle for helpful discussions. This work was supported by a U.S. Department of Energy Quantum Information Science Enabled Discovery (QuantISED) grant for High Energy Physics (KA2401032) and carried out using facilities provided by the U.S. Department of Homeland Security, Domestic Nuclear Detection Office at the Lawrence Berkeley National Laboratory under UC-DOE Contract No. DE-AC02-05CH11231.

# Appendix A

Appendix table I lists the sample numbers, $L_A$, $N_e$, $N_{Si}$, $N_B$, and the decay times and fractions of the exponential components that fit the emission band A decay curves for the 29 samples. Good fits require two exponential components plus a slow component. Emission band A has the lowest luminosity and shortest decay times of the four emission bands.

Appendix table I. Sample numbers, $L_A$, $N_e$, $N_{Si}$, $N_B$, decay times (ns) and fractions for emission band A, sorted by $N_e$

| Sample | $L_A$ | $N_e$ | $N_{Si}$ | $N_B$ | $\tau_1$ | $f_1$ | $\tau_2$ | $f_2$ | $f_{slow}$ |
|---|---|---|---|---|---|---|---|---|---|
| 13357 | 0.22 | 3.8x10⁻⁷ | 0.48 | 1.13 | 54.0 | 0.09 | 655.3 | 0.29 | 0.62 |
| 13358 | 2.07 | 0.02 | 0.18 | 2.96 | 74.2 | 0.08 | 920.9 | 0.28 | 0.64 |
| 13707 | 1.77 | 0.23 | 1.25 | 35.6 | 76.5 | 0.35 | 603.0 | 0.47 | 0.18 |
| 13708 | 0.85 | 0.52 | 2.7 | 38.5 | 49.7 | 0.49 | 354.4 | 0.43 | 0.08 |
| 13359 | 2.89 | 0.53 | 4.3 | 47.4 | 18.5 | 0.67 | 90.4 | 0.30 | 0.03 |
| 13721 | 0.57 | 1.18 | 2.2 | 16.9 | 24.5 | 0.58 | 147.0 | 0.37 | 0.05 |
| 13364 | 0.52 | 1.38 | 5.0 | 14.5 | 11.8 | 0.53 | 69.6 | 0.35 | 0.12 |
| 13365 | 2.24 | 2.24 | 3.5 | 14.5 | 12.7 | 0.78 | 64.9 | 0.19 | 0.03 |
| 13719 | 1.17 | 2.68 | 4.2 | 21.0 | 16.2 | 0.66 | 85.5 | 0.30 | 0.04 |
| 13363 | 0.13 | 4.24 | 4.5 | 6.22 | 2.6 | 0.37 | 96.2 | 0.15 | 0.48 |
| 13704.1 | 2.93 | 4.3 | 5.0 | 13.0 | 10.0 | 0.78 | 48.7 | 0.20 | 0.02 |
| 13701 | 18.1 | 4.6 | 9.2 | 54.6 | 10.1 | 0.80 | 44.5 | 0.18 | 0.01 |
| 13702 | 7.91 | 5.1 | 7.8 | 44.5 | 9.2 | 0.83 | 49.2 | 0.16 | 0.01 |
| 40020 | 3.55 | 5.2 | 10.3 | 44.5 | 8.3 | 0.77 | 41.3 | 0.21 | 0.02 |
| 13316 | 1.32 | 5.5 | 10.2 | 28.7 | 8.7 | 0.73 | 46.7 | 0.22 | 0.05 |
| 13352 | 1.42 | 5.5 | 8.8 | 32.6 | 9.0 | 0.70 | 46.4 | 0.24 | 0.05 |
| 13353 | 0.95 | 5.5 | 8.2 | 27.3 | 9.2 | 0.75 | 51.5 | 0.21 | 0.03 |
| 13354 | 1.02 | 5.5 | 8.8 | 37.1 | 9.4 | 0.72 | 61.2 | 0.17 | 0.11 |
| 40018 | 2.59 | 5.6 | 8.4 | 56.3 | 8.9 | 0.81 | 49.7 | 0.18 | 0.01 |
| 13705.2 | 5.96 | 5.8 | 7.0 | 19.3 | 9.3 | 0.82 | 48.1 | 0.17 | 0.01 |
| 13330 | 1.29 | 6 | 9.5 | 15.4 | 6.4 | 0.79 | 38.4 | 0.18 | 0.03 |
| 13706 | 1.96 | 6 | 17.1 | 53.3 | 5.1 | 0.81 | 27.8 | 0.17 | 0.02 |
| 40019 | 3.70 | 6.8 | 9.8 | 32.6 | 7.6 | 0.81 | 40.4 | 0.17 | 0.02 |
| 13703 | 1.22 | 7 | 8.9 | 41.5 | 7.8 | 0.75 | 43.1 | 0.22 | 0.03 |
| 40017 | 4.43 | 7.6 | 9.7 | 32.6 | 7.2 | 0.85 | 38.8 | 0.14 | 0.01 |
| 40022 | 1.23 | 7.6 | 11.4 | 38.5 | 6.8 | 0.75 | 38.0 | 0.23 | 0.02 |
| 40021 | 2.39 | 9 | 20.5 | 32.6 | 4.9 | 0.76 | 23.1 | 0.22 | 0.02 |
| 13332 | 5.45 | 14.4 | 16.7 | 20.2 | 4.6 | 0.87 | 27.2 | 0.12 | 0.01 |
| 13333 | 5.98 | 20.3 | 22.8 | 23.4 | 4.1 | 0.88 | 23.6 | 0.10 | 0.02 |

Appendix table II lists the sample numbers, $L_B$, $N_e$, $N_{Si}$, $N_B$, and the decay times and fractions of the exponential components that fit the emission band B decay curves for the 29 samples. Good fits for samples with $N_e > 0.23$ require two exponential components plus a slow component. For samples with $N_e \leq 0.23$ a single exponential component plus a slow component is sufficient.

Appendix table II. Sample numbers, $L_B$, $N_e$, $N_{Si}$, $N_B$, and decay times (ns) and fractions for emission band B, sorted by $N_e$

| Sample | $L_B$ | $N_e$ | $N_{Si}$ | $N_B$ | $\tau_1$ | $f_1$ | $\tau_2$ | $f_2$ | $f_{slow}$ |
|---|---|---|---|---|---|---|---|---|---|
| 13357 | 0.006 | $3.8 \times 10^{-7}$ | 0.48 | 1.13 | 732 | 0.50 | | | 0.50 |
| 13358 | 0.20 | 0.02 | 0.18 | 2.96 | 457 | 0.13 | | | 0.87 |
| 13707 | 49.1 | 0.23 | 1.25 | 35.6 | | | 2113 | 0.20 | 0.80 |
| 13708 | 26.0 | 0.52 | 2.7 | 38.5 | 354 | 0.02 | 1677 | 0.43 | 0.54 |
| 13359 | 94.9 | 0.53 | 4.3 | 47.4 | 408 | 0.23 | 1390 | 0.61 | 0.16 |
| 13721 | 0.36 | 1.18 | 2.2 | 16.9 | 403 | 0.06 | 1564 | 0.60 | 0.33 |
| 13364 | 4.24 | 1.38 | 5.0 | 14.5 | 316 | 0.10 | 1255 | 0.70 | 0.19 |
| 13365 | 8.40 | 2.24 | 3.5 | 14.5 | 361 | 0.27 | 1167 | 0.65 | 0.08 |
| 13719 | 0.73 | 2.68 | 4.2 | 21.0 | 350 | 0.28 | 1253 | 0.61 | 0.11 |
| 13363 | 13.7 | 4.24 | 4.5 | 6.22 | 245 | 0.42 | 704 | 0.56 | 0.02 |
| 13704.1 | 22.1 | 4.3 | 5.0 | 13.0 | 254 | 0.52 | 821 | 0.46 | 0.02 |
| 13701 | 69.0 | 4.6 | 9.2 | 54.6 | 239 | 0.51 | 763 | 0.46 | 0.02 |
| 13702 | 26.3 | 5.1 | 7.8 | 44.5 | 208 | 0.58 | 648 | 0.41 | 0.01 |
| 40020 | 27.8 | 5.2 | 10.3 | 44.5 | 173 | 0.60 | 541 | 0.38 | 0.01 |
| 13316 | 26.4 | 5.5 | 10.2 | 28.7 | 202 | 0.52 | 653 | 0.46 | 0.02 |
| 13352 | 31.4 | 5.5 | 8.8 | 32.6 | 213 | 0.51 | 707 | 0.47 | 0.02 |
| 13353 | 24.8 | 5.5 | 8.2 | 27.3 | 205 | 0.51 | 674 | 0.47 | 0.02 |
| 13354 | 24.6 | 5.5 | 8.8 | 37.1 | 197 | 0.49 | 644 | 0.49 | 0.02 |
| 40018 | 14.1 | 5.6 | 8.4 | 56.3 | 203 | 0.59 | 629 | 0.40 | 0.01 |
| 13705.2 | 33.3 | 5.8 | 7.0 | 19.3 | 211 | 0.60 | 665 | 0.39 | 0.01 |
| 13330 | 8.74 | 6 | 9.5 | 15.4 | 156 | 0.57 | 468 | 0.42 | 0.01 |
| 13706 | 3.95 | 6 | 17.1 | 53.3 | 92 | 0.62 | 252 | 0.37 | 0.01 |
| 40019 | 18.4 | 6.8 | 9.8 | 32.6 | 160 | 0.66 | 479 | 0.34 | 0.01 |
| 13703 | 10.4 | 7 | 8.9 | 41.5 | 172 | 0.53 | 505 | 0.46 | 0.01 |
| 40017 | 11.8 | 7.6 | 9.7 | 32.6 | 157 | 0.67 | 460 | 0.32 | 0.01 |
| 40022 | 9.89 | 7.6 | 11.4 | 38.5 | 141 | 0.63 | 442 | 0.37 | 0.01 |
| 40021 | 5.16 | 9 | 20.5 | 32.6 | 81 | 0.67 | 244 | 0.31 | 0.01 |
| 13332 | 5.46 | 14.4 | 16.7 | 20.2 | 58 | 0.59 | 155 | 0.40 | 0.01 |
| 13333 | 3.70 | 20.3 | 22.8 | 23.4 | 42 | 0.61 | 116 | 0.37 | 0.02 |

Appendix table III lists the sample numbers, $L_C$, $N_e$, $N_{Si}$, $N_B$, and the decay times and fractions of the exponential components that fit the emission band C decay curves for the 29 samples. Good fits for samples with $N_e > 0.02$ require two exponential components plus a slow component. For samples with $N_e \leq 0.02$ a single exponential component plus a slow component is sufficient.

Appendix table III. Sample numbers, $L_C$, $N_e$, $N_{Si}$, $N_B$, and decay times (ns) and fractions for emission band C, sorted by $N_e$

| Sample | $L_C$ | $N_e$ | $N_{Si}$ | $N_B$ | $\tau_1$ | $f_1$ | $\tau_2$ | $f_2$ | $f_{slow}$ |
|---|---|---|---|---|---|---|---|---|---|
| 13357 | 0.008 | 3.8x10$^{-7}$ | 0.48 | 1.13 | 294 | 0.15 | | | 0.85 |
| 13358 | 0.71 | 0.02 | 0.18 | 2.96 | 66 | 0.01 | | | 0.99 |
| 13707 | 4.86 | 0.23 | 1.25 | 35.6 | 44 | 0.002 | 2314 | 0.11 | 0.89 |
| 13708 | 4.61 | 0.52 | 2.7 | 38.5 | 91 | 0.003 | 1992 | 0.21 | 0.79 |
| 13359 | 4.92 | 0.53 | 4.3 | 47.4 | 278 | 0.056 | 1451 | 0.55 | 0.40 |
| 13721 | 8.96 | 1.18 | 2.2 | 16.9 | 40 | 0.006 | 1975 | 0.37 | 0.62 |
| 13364 | 7.05 | 1.38 | 5.0 | 14.5 | 30 | 0.006 | 1888 | 0.48 | 0.52 |
| 13365 | 4.89 | 2.24 | 3.5 | 14.5 | 19 | 0.005 | 1547 | 0.65 | 0.34 |
| 13719 | 26.9 | 2.68 | 4.2 | 21.0 | 229 | 0.02 | 1633 | 0.57 | 0.40 |
| 13363 | 5.82 | 4.24 | 4.5 | 6.22 | 282 | 0.11 | 1173 | 0.72 | 0.17 |
| 13704.1 | 8.77 | 4.3 | 5.0 | 13.0 | 328 | 0.13 | 1368 | 0.71 | 0.16 |
| 13701 | 5.44 | 4.6 | 9.2 | 54.6 | 270 | 0.14 | 1044 | 0.68 | 0.18 |
| 13702 | 3.21 | 5.1 | 7.8 | 44.5 | 255 | 0.18 | 1026 | 0.69 | 0.13 |
| 40020 | 7.18 | 5.2 | 10.3 | 44.5 | 263 | 0.24 | 1005 | 0.66 | 0.10 |
| 13316 | 25.8 | 5.5 | 10.2 | 28.7 | 338 | 0.17 | 1310 | 0.70 | 0.14 |
| 13352 | 37.4 | 5.5 | 8.8 | 32.6 | 346 | 0.15 | 1365 | 0.69 | 0.15 |
| 13353 | 27.5 | 5.5 | 8.2 | 27.3 | 335 | 0.15 | 1340 | 0.70 | 0.15 |
| 13354 | 31.1 | 5.5 | 8.8 | 37.1 | 324 | 0.15 | 1317 | 0.70 | 0.15 |
| 40018 | 2.73 | 5.6 | 8.4 | 56.3 | 312 | 0.24 | 1182 | 0.66 | 0.097 |
| 13705.2 | 5.28 | 5.8 | 7.0 | 19.3 | 315 | 0.23 | 1171 | 0.67 | 0.103 |
| 13330 | 29.4 | 6 | 9.5 | 15.4 | 300 | 0.27 | 1118 | 0.66 | 0.065 |
| 13706 | 4.46 | 6 | 17.1 | 53.3 | 176 | 0.45 | 638 | 0.52 | 0.026 |
| 40019 | 4.74 | 6.8 | 9.8 | 32.6 | 258 | 0.31 | 920 | 0.62 | 0.069 |
| 13703 | 3.61 | 7 | 8.9 | 41.5 | 248 | 0.22 | 957 | 0.70 | 0.086 |
| 40017 | 3.91 | 7.6 | 9.7 | 32.6 | 263 | 0.32 | 914 | 0.63 | 0.050 |
| 40022 | 8.39 | 7.6 | 11.4 | 38.5 | 276 | 0.34 | 1039 | 0.60 | 0.057 |
| 40021 | 8.81 | 9 | 20.5 | 32.6 | 152 | 0.52 | 571 | 0.46 | 0.020 |
| 13332 | 19.7 | 14.4 | 16.7 | 20.2 | 113 | 0.57 | 409 | 0.42 | 0.013 |
| 13333 | 2.79 | 20.3 | 22.8 | 23.4 | 76 | 0.60 | 268 | 0.39 | 0.014 |

Appendix table IV lists the sample numbers, $L_D$, $N_e$, $N_{Si}$, $N_B$, and the decay times and fractions of the exponential components that fit the emission band D decay curves for the 29 samples. Good fits require one exponential components plus a slow component. Band D has the slowest decay time of all the emission bands.

Appendix table IV. Sample numbers, $L_D$, $N_e$, $N_{Si}$, $N_B$, and decay times (ns) and fractions for emission band D, sorted by $N_e$

| Sample | $L_D$ | $N_e$ | $N_{Si}$ | $N_B$ | $\tau_1$ | $f_1$ | $f_{slow}$ |
|---|---|---|---|---|---|---|---|
| 13357 | 0.06 | 3.8x10$^{-7}$ | 0.48 | 1.13 | 603 | 0.35 | 0.65 |
| 13358 | 2.04 | 0.02 | 0.18 | 2.96 | 58 | 0.007 | 0.99 |
| 13707 | 24.8 | 0.23 | 1.25 | 35.6 | 3956 | 0.009 | 0.99 |
| 13708 | 25.9 | 0.52 | 2.7 | 38.5 | 3156 | 0.025 | 0.98 |
| 13359 | 15.6 | 0.53 | 4.3 | 47.4 | 3483 | 0.086 | 0.91 |
| 13721 | 51.7 | 1.18 | 2.2 | 16.9 | 3152 | 0.057 | 0.94 |
| 13364 | 67.7 | 1.38 | 5.0 | 14.5 | 2899 | 0.096 | 0.90 |
| 13365 | 24.0 | 2.24 | 3.5 | 14.5 | 2687 | 0.19 | 0.81 |
| 13719 | 43.9 | 2.68 | 4.2 | 21.0 | 3304 | 0.17 | 0.83 |
| 13363 | 6.78 | 4.24 | 4.5 | 6.22 | 2204 | 0.49 | 0.51 |
| 13704.1 | 10.1 | 4.3 | 5.0 | 13.0 | 2331 | 0.42 | 0.58 |
| 13701 | 3.60 | 4.6 | 9.2 | 54.6 | 1790 | 0.42 | 0.58 |
| 13702 | 1.86 | 5.1 | 7.8 | 44.5 | 1538 | 0.48 | 0.52 |
| 40020 | 3.96 | 5.2 | 10.3 | 44.5 | 1564 | 0.64 | 0.36 |
| 13316 | 29.6 | 5.5 | 10.2 | 28.7 | 2168 | 0.51 | 0.49 |
| 13352 | 47.3 | 5.5 | 8.8 | 32.6 | 2253 | 0.47 | 0.53 |
| 13353 | 31.0 | 5.5 | 8.2 | 27.3 | 2220 | 0.49 | 0.51 |
| 13354 | 36.7 | 5.5 | 8.8 | 37.1 | 2238 | 0.50 | 0.50 |
| 40018 | 1.62 | 5.6 | 8.4 | 56.3 | 1852 | 0.58 | 0.42 |
| 13705.2 | 3.94 | 5.8 | 7.0 | 19.3 | 1980 | 0.59 | 0.41 |
| 13330 | 16.5 | 6 | 9.5 | 15.4 | 1605 | 0.65 | 0.35 |
| 13706 | 0.96 | 6 | 17.1 | 53.3 | 689 | 0.76 | 0.24 |
| 40019 | 1.74 | 6.8 | 9.8 | 32.6 | 1152 | 0.64 | 0.36 |
| 13703 | 1.67 | 7 | 8.9 | 41.5 | 1327 | 0.62 | 0.38 |
| 40017 | 2.18 | 7.6 | 9.7 | 32.6 | 1266 | 0.70 | 0.30 |
| 40022 | 3.38 | 7.6 | 11.4 | 38.5 | 1363 | 0.71 | 0.29 |
| 40021 | 0.96 | 9 | 20.5 | 32.6 | 567 | 0.80 | 0.20 |
| 13332 | 2.04 | 14.4 | 16.7 | 20.2 | 412 | 0.89 | 0.11 |
| 13333 | 0.87 | 20.3 | 22.8 | 23.4 | 233 | 0.84 | 0.16 |

Appendix figure 1 is a plot of the seven decay times as a function of $N_B$. There is no systematic correlation, consistent with the idea that $N_e$ is the primary controller of the decay times.

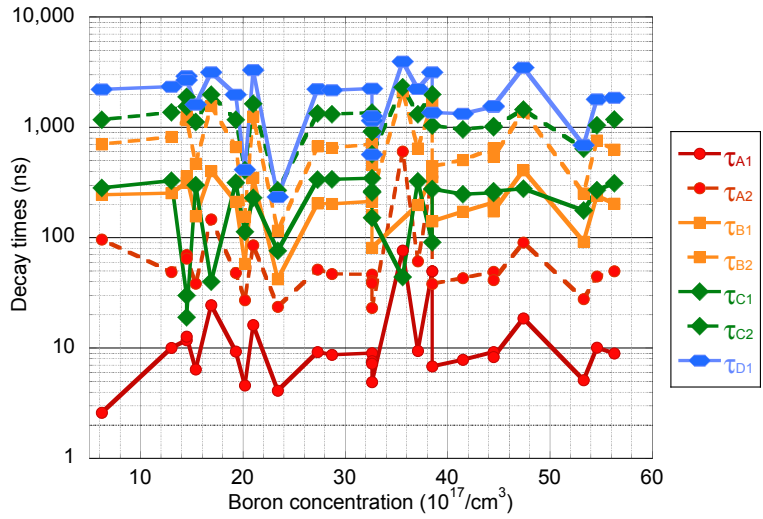

Appendix figure 1. Decay times for the exponential components listed in tables I to IV vs. boron concentration. Solid lines are for components A1, B1, C1, and D1. Dashed lines are for components A2, B2, and C2. There is no apparent systematic correlation between decay times and boron concentration.

Appendix tables V to VIII list the luminosities (photons/keV), decay times (ns), and fractions of the four samples filtered for the four emission bands as a function of temperature.

Appendix table V. Luminosities, exponential decay times and fractions vs. temperature for emission band A recorded from sample 13701 using narrow band filter NB850 and short pass filter SP1200. Exponential component fits are not reported above 50K due to low statistics.

| Temperature (K) | Luminosity | $\tau_1$ (ns) | $f_1$ | $\tau_2$ (ns) | $f_2$ | $f_{slow}$ |
|---|---|---|---|---|---|---|
| 10 | 18.12 | 9.0 | 0.70 | 36.0 | 0.28 | 0.01 |
| 25 | 13.84 | 9.6 | 0.77 | 44.1 | 0.22 | 0.02 |
| 50 | 8.99 | 7.9 | 0.66 | 37.2 | 0.32 | 0.02 |
| 100 | 0.37 | | | | | |
| 150 | 0.16 | | | | | |
| 200 | 0.15 | | | | | |
| 250 | 0.08 | | | | | |
| 293 | 0.06 | | | | | |

Appendix table VI. Luminosities, exponential decay times and fractions vs. temperature for emission band B recorded from sample 13359 using narrow band filter NB925 and short pass filter SP1200. Exponential component fits are not reported above 150K due to low statistics.

| Temperature (K) | Luminosity | $\tau_1$ (ns) | $f_1$ | $\tau_2$ (ns) | $f_2$ | $f_{slow}$ |
|---|---|---|---|---|---|---|
| 10 | 94.87 | 1394 | 0.62 | 409 | 0.20 | 0.18 |
| 25 | 89.65 | 1404 | 0.63 | 412 | 0.20 | 0.18 |
| 50 | 77.21 | 1398 | 0.64 | 411 | 0.19 | 0.17 |
| 100 | 61.47 | 1368 | 0.71 | 354 | 0.13 | 0.16 |
| 150 | 27.15 | 1361 | 0.79 | 209 | 0.09 | 0.12 |
| 200 | 2.81 | | | | | |
| 250 | 0.30 | | | | | |
| 293 | 0.21 | | | | | |

Appendix table VII. Luminosities, exponential decay times and fractions vs. temperature for emission band C recorded from sample 13330 using narrow band filter NB1075 and short pass filter SP1200. Exponential component fits are not reported above 150K due to low statistics.

| Temperature (K) | Luminosity | $\tau_1$ (ns) | $f_1$ | $\tau_2$ (ns) | $f_2$ | $f_{slow}$ |
|---|---|---|---|---|---|---|
| 10 | 29.41 | 1101 | 0.66 | 280 | 0.28 | 0.07 |
| 25 | 27.39 | 1068 | 0.67 | 268 | 0.26 | 0.07 |
| 50 | 24.34 | 1060 | 0.67 | 268 | 0.26 | 0.07 |
| 100 | 17.24 | 1022 | 0.68 | 260 | 0.25 | 0.07 |
| 150 | 11.76 | 934 | 0.71 | 243 | 0.23 | 0.06 |
| 200 | 3.43 | | | | | |
| 250 | 0.80 | | | | | |
| 293 | 0.58 | | | | | |

Appendix table VIII. Luminosities, exponential decay times and fractions vs. temperature for emission band D recorded from sample 13364 using narrow band filter NB1350. This is the only band with measurable luminosity at room temperature.

| Temperature (K) | Luminosity | $\tau_1$ (ns) | $f_1$ | $f_{slow}$ |
|---|---|---|---|---|
| 10 | 67.74 | 3036 | 0.10 | 0.90 |
| 25 | 63.96 | 2899 | 0.10 | 0.90 |
| 50 | 55.27 | 3004 | 0.11 | 0.89 |
| 100 | 41.04 | 2924 | 0.11 | 0.89 |
| 150 | 30.58 | 2804 | 0.13 | 0.87 |
| 200 | 24.68 | 2873 | 0.14 | 0.86 |
| 250 | 18.99 | 2662 | 0.15 | 0.85 |
| 293 | 13.35 | 2447 | 0.17 | 0.83 |